\documentstyle[12pt,a4wide,epsf]{article}
\newlength{\absize}
\renewcommand{\arraystretch}{0.5}
\newenvironment{describe}{\small
\begin{minipage}[c]{40mm}\setlength{\baselineskip}{0ex}
\raggedright\vspace{4pt}}{\vspace{4pt}\end{minipage}}
\newenvironment{describea}{\small
\begin{minipage}[c]{25mm}\setlength{\baselineskip}{0ex}
\raggedright\vspace{4pt}}{\vspace{4pt}\end{minipage}}
\begin{document}
\thispagestyle{empty}
\pagestyle{empty}
\renewcommand{\thefootnote}{\fnsymbol{footnote}}
\newcommand{\starttext}{\newpage\normalsize
  \pagestyle{plain}
  \setlength{\baselineskip}{4ex}\par
  \setcounter{footnote}{0}
  \renewcommand{\thefootnote}{\arabic{footnote}}}

\renewcommand{\theequation}{\thesection.\arabic{equation}}
\newcommand{\preprint}[1]{{\tt
  \begin{flushright}
    \setlength{\baselineskip}{3ex} #1
  \end{flushright}}}
\renewcommand{\title}[1]{%
  \begin{center}
    \LARGE #1
  \end{center}\par}
\renewcommand{\author}[1]{%
  \vspace{2ex}
  {\Large
   \begin{center}
     \setlength{\baselineskip}{3ex} #1 \par
   \end{center}}}
\renewcommand{\thanks}[1]{\footnote{#1}}
\renewcommand{\abstract}[1]{%
  \vspace{2ex}
  \normalsize
  \begin{center}
    \centerline{\bf Abstract}\par
    \vspace{2ex}
    \parbox{\absize}{#1\setlength{\baselineskip}{2.5ex}\par}
  \end{center}}

\setlength{\parindent}{3em}
\setlength{\footnotesep}{.6\baselineskip}
\newcommand{\myfoot}[1]{%
  \footnote{\setlength{\baselineskip}{.75\baselineskip}#1}}
\renewcommand{\thepage}{\arabic{page}}
\setcounter{bottomnumber}{2}
\setcounter{topnumber}{3}
\setcounter{totalnumber}{4}
\newcommand{\figsize}{}
\renewcommand{\bottomfraction}{1}
\renewcommand{\topfraction}{1}
\renewcommand{\textfraction}{0}
\newcommand{\beq}{\begin{equation}}
\newcommand{\eeq}{\end{equation}}
\newcommand{\beqa}{\begin{eqnarray}}
\newcommand{\eeqa}{\end{eqnarray}}
\newcommand{\nn}{\nonumber}

\newcommand{\dd}{\mbox{{\rm d}}}

\newcommand{\Dis}[1]{$\displaystyle #1$}
\newcommand{\Disp}[1]{{\displaystyle #1}}

\newcommand{\hc}{\mbox{{\rm h.c.}}}
\newcommand{\MNS}{M_2}
\newcommand{\mlN}{M_1}
\newcommand{\Mu}{m_{{ u}}}
\newcommand{\Au}{A_{{ u}}}
\newcommand{\Md}{m_{{ d}}}
\newcommand{\Ad}{A_{{ d}}}
\newcommand{\Ab}{A_{{ b}}}
\newcommand{\At}{A_{{ t}}}
\newcommand{\suq}{\tilde{{ u}}}
\newcommand{\Tsp}{\mbox{\scriptsize T}}
\newcommand{\eps}{\epsilon}
\newcommand{\sdq}{\tilde{{ d}}}
\newcommand{\gluino}{\tilde g}
\newcommand{\squark}{\tilde q}
\newcommand{\thW}{\theta_{\rm W}}
\newcommand{\mW}{m_{{ W}}}
\newcommand{\MsQU}{\wtilde{3}{0.8}{M}_{\hspace*{-1mm}U}}
\newcommand{\MsQT}{\wtilde{3}{0.8}{M}_{\hspace*{-1mm}T}}
\newcommand{\Msu}{\wtilde{3}{0.2}{m}_{U}}
\newcommand{\Msd}{\wtilde{3}{0.2}{m}_{\!D}}
\newcommand{\Msq}{m_{\tilde{q}}}
\newcommand{\MsT}{\wtilde{3}{0.2}{m}_{T}}
\newcommand{\Ms}{\wtilde{3}{0.8}{m}}
\newcommand{\wtilde}[3]{\settowidth{\ltT}{\Dis{#3}}
\makebox[\ltT]{$\rule{#2\mmh}{0mm}
\widetilde{\makebox[#1\mm]{\Dis{#3\rule{#2\mm}{0mm}}}}$}}

\newlength{\ltT}
\newlength{\mmh}
\setlength{\mmh}{0.5mm}
\newlength{\mm}
\setlength{\mm}{1mm}

\def\Im{\mbox{\rm Im\ }}
\def\Re{\mbox{\rm Re\ }}
\def\fourth{\textstyle{1\over4}}
\def\gsim{\mathrel{\rlap{\raise 1.5pt \hbox{$>$}}\lower 3.5pt
\hbox{$\sim$}}}
\def\lsim{\mathrel{\rlap{\raise 1.5pt \hbox{$<$}}\lower 3.5pt
\hbox{$\sim$}}}
\def\GeV{{\rm GeV}}
\def\TeV{{\rm TeV}}
\def\Order{{\cal O}}
\catcode`@=11
\def\citer{\@ifnextchar [{\@tempswatrue\@citexr}{\@tempswafalse\@citexr[]}}

%

\def\@citexr[#1]#2{\if@filesw\immediate\write\@auxout{\string\citation{#2}}\fi
  \def\@citea{}\@cite{\@for\@citeb:=#2\do
    {\@citea\def\@citea{--\penalty\@m}\@ifundefined
       {b@\@citeb}{{\bf ?}\@warning
       {Citation `\@citeb' on page \thepage \space undefined}}%
\hbox{\csname b@\@citeb\endcsname}}}{#1}}
\catcode`@=12
%
\def\slash#1{#1 \hskip -0.5em /}
\def\check{{\it (check this!)\ }}
%
\def\Month{\ifcase\month\or
January\or February\or March\or April\or May\or June\or
July\or August\or September\or October\or November\or December\fi}
\def\slash#1{#1 \hskip -0.5em /}
%

\begin{flushright}
NORDITA -- 95/48 P\\
hep-ph/9506455
\end{flushright}

\vfill
\title{Production and Two-photon Decay of
the MSSM Scalar Higgs Bosons at the LHC}

\vfill
\author{B. Kileng${}^a$, P. Osland${}^b$, P.N. Pandita${}^{a,c}$ \\ \hfil\\
    ${}^{a}$ NORDITA, Blegdamsvej 17, DK-2100 Copenhagen \O, Denmark \\
    ${}^{b}$ Department of Physics, University of Bergen,
    All\'egt.~55, N-5007 Bergen, Norway\\
    ${}^{c}$ North Eastern Hill University, Laitumkhrah, Shillong 793003,
    India\thanks{Permanent address}}
\date{}

\vfill
\abstract{
We consider the production and two-photon decay of the $CP$-even
Higgs bosons ($h^0$ and $H^0$) of the Minimal Supersymmetric
Standard Model (MSSM) at the Large Hadron Collider.
We study in detail the dependence of the cross section
on various parameters of the MSSM, especially the dependence
on the mixing effects in the squark sector due to the Higgs
bilinear parameter $\mu$ and the soft supersymmetry breaking
parameter $A$.
We find that the cross section for the production of these
Higgs bosons has a significant dependence on the parameters
which determine the chiral mixing in the squark sector.
The cross section times the two-photon branching ratio
of $h^0$ is of the order of 15--25~fb in much
of the parameter space that remains after imposing the
present  experimental constraints.
For the $H^0$ the two-photon branching ratio is only significant
if the $H^0$ is light, but then the cross section times the
branching ratio may exceed 200~fb.
The QCD corrections due to quark loop contributions
are known to increase the cross section by 50\%.
We find the dependence of the cross section on the gluon
distribution function used to be rather insignificant.
}

\vfill

\starttext

\section{Introduction}
\label{sec:intro}
\setcounter{equation}{0}
It is well known that in the Minimal Supersymmetric Standard Model (MSSM)
\cite{MSSM}, two Higgs doublets with opposite hypercharge are
required in order to preserve supersymmetry.
The physical Higgs boson spectrum in the MSSM consists of
two $CP$-even neutral bosons $h^0$ and $H^0$, a $CP$-odd neutral
boson $A^0$ and a pair of charged Higgs bosons $H^\pm$.
The most important production mechanism for the neutral SUSY Higgs
particles at the Large Hadron Collider (LHC) is the gluon
fusion mechanism, $pp\to gg\to h^0$, $H^0$, $A^0$
\cite{GGMN}
and the Higgs radiation off top and bottom quarks
\cite{Kunsztetal}.
Except for the small range in the parameter space where the
heavy neutral Higgs $H^0$ decays into a pair of $Z$ bosons,
the rare $\gamma\gamma$ decay mode, apart from $\tau\tau$ decays,
is a promising mode to detect the neutral Higgs particles,
especially if $b$ quark decays cannot be separated
from the QCD background\footnote{It should be kept in mind that if SUSY
particles are light, decays to squarks and charginos,
$h^0\to\tilde q\tilde q$, $h^0\to\tilde\chi^+\tilde\chi^-$,
could be seen at LEP2.}.

This process was studied several years ago \cite{KunsztZ},
and it was concluded that the lightest Higgs could be detected
in this mode for sufficiently large values of the mass of the
pseudoscalar Higgs boson $m_A\gg m_Z$.
Similarly, the $\gamma\gamma$ channel is important
for the discovery of $H^0$ for $50~\GeV\le m_A\le 150~\GeV$.
Related studies have been presented in ref.\ \cite{Baer}.
While most of the emphasis in these earlier works was on the SSC, the
discussion of the dependence of the cross section on various parameters
is relevant.

In this paper we present a fresh study of the hadronic production
and subsequent two-photon decay of the $CP$-even Higgs bosons
($h^0$ and $H^0$) of the MSSM,
valid for the LHC energy of $\sqrt{s}=14~\TeV$,
and using gluon distribution functions based on recent HERA
data \cite{Plothow},
to reassess the feasibility of observing the $CP$-even Higgs
bosons in this mode.
As mentioned earlier, the gluon fusion mechanism is the dominant
production mechanism of SUSY Higgs bosons in high-energy
$pp$ collisions throughout the entire Higgs mass range.
We study the cross section for the production of the $h^0$
and $H^0$, and their decays, taking into account all
the parameters of the model. In particular, we take into account
the mixing in the squark sector, the chiral mixing,
which also affects the Higgs boson masses
through appreciable radiative corrections.
This was previously shown to lead to large corrections to the rates
\cite{Kileng}.

In the calculation of the production of the Higgs through
gluon-gluon fusion, we include in the triangle graph loop
all the squarks, as well as $b$ and $t$ quarks, the lightest quarks
having a negligible coupling to the $h^0$.
On the other hand, in the calculation of decay of the Higgs to two
photons, we include in addition to the above, all the sleptons,
$W^\pm$, charginos and the charged Higgs boson.

The Minimal Supersymmetric Model contains several
soft supersymmetry-breaking terms.
We write the relevant soft terms in the Lagrangian
as follows \cite{GirGri}
\begin{eqnarray}
\label{EQU:Lagrangefour}
{\cal L}_{\mbox{{\scriptsize Soft}}}
& = & \Biggl\{ \frac{g\Md\Ad}{\sqrt{2}\;\mW\cos\beta}Q^{\Tsp}\eps H_{1}\sdq^{R}
    - \frac{g\Mu\Au}{\sqrt{2}\;\mW\sin\beta}Q^{\Tsp}\eps H_{2}\suq^{R}
    + \hc \Biggr\} \nonumber \\
& & - \MsQU^{2}Q^{\dagger}Q - \Msu^{2}\suq^{R\dagger}\suq^{R}
    - \Msd^{2}\sdq^{R\dagger}\sdq^{R}
    - M_{\!H_{1}}^{2} H_{1}^{\dagger}H_{1}
    - M_{\!H_{2}}^{2} H_{2}^{\dagger}H_{2} \nn \\*
& & + \frac{\mlN}{2}\left\{\lambda\lambda +\bar{\lambda}\;
      \bar{\lambda}\right\}
    + \frac{\MNS}{2} \sum_{k=1}^{3} \left\{\Lambda^{k}\Lambda^{k}
    + \bar{\Lambda}^{k}\bar{\Lambda}^{k}\right\}.
\end{eqnarray}
Subscripts $u$ (or $U$) and $d$ (or $D$) refer generically to up and
down-type quarks.
The Higgs production cross
section and the two-photon decay rate depend significantly
on several of these parameters.

Even without chiral mixing, two basic mass scales are required,
those of squark and gaugino masses.
The squark masses are determined by an SU(2)-doublet mass parameter,
together with two SU(2)-singlet mass parameters,
denoted in eq.~(\ref{EQU:Lagrangefour})
by $\MsQU$, $\Msu$ and $\Msd$, respectively.
As is often done, we consider the case when all the SUSY-breaking
squark masses are taken equal to a common mass parameter,
which we denote by $\Ms$ ($=\MsQU=\Msu=\Msd$).
We shall consider the situation where this parameter is chosen
to be 150~GeV for the first two generations, and vary it over
the values 150, 500 and 1000~GeV for the third generation.
(Most of the plots presented will be for $\Ms=500~\GeV$.)
The physical squark masses will in general be different
from these values, but they determine the relevant
orders of magnitude.

The contributions from the squark and Higgs sectors depend on
the relative sign between the $A$ and $\mu$ parameters, but not on
their overall signs. I.e., we can choose $A$ positive, but then we
must study both negative and positive values of $\mu$ in order to
cover all of parameter space.
The Higgs sector depends on $A$ and $\mu$ through the
radiative corrections.

The contribution from the chargino sector depends on the relative sign
between $\mu$ and $\MNS$. The chargino contribution
is independent of $A$. Thus, the $h^{0}\to\gamma\gamma$ decay rate
is independent of the the over-all signs of $A$, $\mu$ and $\MNS$,
but depends on all the relative signs.
To be more specific, we may choose $\MNS$ positive,
but then $A$ and $\mu$ must both be allowed to take on
negative and positive values in order to cover the full parameter space.
In most of the parameter space though, the dependence on
the chargino mass (and therefore also on the sign of $\MNS$) is
rather weak.  In these regions it suffices to consider $A$ positive.

The signs of the off-diagonal terms in the squark mass matrices
are given by the definition of $A$ and $\mu$, and also
by the definition of the fermion masses.
The sign chosen for the fermion mass terms show up
in some of the couplings, in the fermion propagators
and the spin sums.

Having $\mu$ and $A$ non-zero has implications for the Higgs masses
as well.  These will be presented in Sec.~2, together with constraints
related to other masses,
before we come to the study of the cross sections and decay rates
in Secs.~3 and 4.
\section{Constraints on Parameter Space}
\label{sec:limit}
\setcounter{equation}{0}
As discussed in the Introduction, there are various contributions to the
production and decay of the lightest Higgs boson at the LHC,
and, therefore, it is necessary to have a description of the full parameter
space and the theoretical and experimental constraints on it
before embarking on the calculation of the cross section and the decay.

At the tree level, the $CP$-even neutral Higgs mass matrix is controlled
by two parameters, which can be chosen to be $m_A$ and $\tan\beta$.
However, there are substantial radiative corrections
\cite{Okada}
to the neutral Higgs masses which depend on,
besides the top quark mass, the
supersymmetric trilinear couplings ($\Au$, $\Ad$), the soft
supersymmetry breaking masses ($\MsQU$, $\Msu$, $\Msd$, etc.),
the bilinear parameter $\mu$ in the superpotential, and
$\tan\beta$ ($=v_2/v_1$, where $v_2$ and $v_1$ are the vacuum
expectation values of the two Higgs doublets of MSSM).
More recent radiative corrections \cite{Carena}
typically reduce the Higgs mass by 10--20~GeV.
However, they are only valid when the squark masses are
of the same order of magnitude, and will therefore not be used
in the present study.
As long as the ``loop particles" are far from threshold for
real production, the cross section does not depend very strongly
on the exact value of the Higgs mass.

We shall assume that all the trilinear couplings
are equal so that
\beq
\Au=\Ad\equiv A,
\eeq
and take the top-quark mass to be 176~GeV
\cite{CDF} in our numerical calculations.
We vary the parameters which enter the neutral $CP$-even Higgs mass
matrix in the following ranges:
\beqa
50~\GeV&\leq& m_A \leq 1000~\GeV, \qquad 1.1\leq\tan\beta\leq 50.0,
\nn \\
\qquad 50~\GeV&\leq&|\mu|\leq1000~\GeV, \qquad
0\leq A\leq 1000~\GeV.
\eeqa

Parts of the $\mu$-$\tan\beta$ plane must be excluded
because of the experimental constraints on
the squark, chargino and $h^0$ masses.
For low values of $\Ms$, the lightest squark tends to be
too light (below the most rigorous experimental bound,
$\sim 44.5~\GeV$ \cite{squarklimit}) or even unphysical
(mass squared negative).
The excluded region of the parameter space is indicated
in fig.~1 for $\Ms=150~\GeV$, $\MNS=200~\GeV$, $m_A=200~\GeV$
and two values of the trilinear coupling $A$.

The allowed region decreases with increasing $A$,
but the dependence on $\MNS$ and $m_A$ is in this region
rather weak.
In order to have acceptable $b$-squarks,
$\mu$ and $\tan\beta$ must lie {\it inside} of the
hyperbola-shaped curves. Similarly, in order to have
acceptable $t$-squarks, the corners at large $|\mu|$ and
small $\tan\beta$ must be excluded.


The chargino masses are, at the tree level, given by the expression
\beqa
m^2_{\chi^\pm}
&=& \frac{1}{2}(\MNS^2+\mu^2)+\mW^2 \nn \\
& & \pm\left[\frac{1}{4}(\MNS^2-\mu^2)^2 +\mW^4\cos^2{2\beta}
+\mW^2(\MNS^2+\mu^2+2\mu\MNS\sin{2\beta})\right]^{1/2}
\eeqa
For $\mu=0$, the above expression simplifies to:
\beq
m^2_{\chi^\pm} = \frac{1}{2}\MNS^2+\mW^2
\pm\left[\frac{1}{4}\MNS^4+\mW^2\cos^2{2\beta}+\mW^2\MNS^2\right]^{1/2}
\eeq
For the case of $\mu=0$, we see that, for $\tan\beta\gg1$,
the lightest chargino becomes massless.
Actually, small values of $\mu$ are unacceptable
for all values of $\tan\beta$.
The lower acceptable value for $|\mu|$ will depend
on $\tan\beta$, but that dependence is rather weak.
The region that is excluded due to the chargino being too light,
increases with decreasing values of $\MNS$.
We note that the radiative corrections to the chargino masses are small
for most of the parameter space \cite{Lahanas}.
We show in fig.~1 the contours in the $\mu$-$\tan\beta$
plane outside of which the chargino has an acceptable mass ($>45~\GeV$)
\cite{chino}.
By the time the LHC starts
operating, one would have searched for charginos with
masses up to 90~GeV at LEP2. Contours relevant for LEP2
are also shown.


For larger values of $\Ms$, there is no problem with squark masses.
However, then the radiative corrections to the Higgs masses
get large, and correspondingly some regions of parameter
space have to be excluded.
This is illustrated in fig.~2, for $\Ms=500~\GeV$.
The corners at large values of $|\mu|$ and $\tan\beta$ must
be excluded since the $h^0$ mass there would fall below
the experimental bound obtained at LEP \cite{LEPdata}.
The extent of these forbidden corners grows rapidly as $m_A$
decreases below $\Order(150~\GeV)$.
They also increase with increasing values of $A$.


The charged Higgs boson mass is given by
\beq
m^2_{H^\pm}=\mW^2+m_A^2+\Delta
\eeq
where $\Delta$ arises due to radiative corrections
and is a complicated function of the parameters of the model
\cite{Brignole}.

The radiative corrections to the charged Higgs mass are not, in general,
as large as in the case of neutral Higgs bosons.
This is due to an approximate global $SU(2)\times SU(2)$ symmetry
\cite{Haber}, valid in the limit of no mixing.
In certain regions of parameter space the radiative corrections can,
however, be large.
This is the case when the trilinear mixing parameter $A$ is
large, $m_A$ is small, and when furthermore $\tan\beta$ is large.
We shall include the effects of non-zero $A$ and $\mu$
in the calculation of the charged Higgs mass.
The present experimental limit of order 40--45~GeV
\cite{chargedH}
is not relevant,
but presumably by the time the LHC starts operating, one will at LEP2
have searched for charged Higgs bosons with mass up to around 90~GeV.
Even this bound does not appreciably restrict the parameter space
as given in figs.~1 and 2.


The neutralino mass matrix depends on four parameters.
These are $\MNS$, $\mlN$, $\mu$ and $\tan\beta$.
However, one can reduce the number of parameters by assuming that
the MSSM is embedded in a grand unified theory so that the
SUSY-breaking gaugino masses are equal to
a common mass at the grand unified scale.
At the electroweak scale, we then have \cite{Inoue}
\beq
\mlN=\frac{5}{3}\tan^2\thW\, \MNS
\eeq
We shall assume this relation throughout in what follows.
The neutralino masses enter the calculation through the total
width of the Higgs boson.
For the gaugino masses, we take $\MNS$ to be 50, 200, or 1000~GeV.
(Most plots will be for $\MNS=200~\GeV$.)
The experimental constraint on the lightest
neutralino mass rules out certain regions of the parameter space
\cite{neutr}, but these depend on several parameters,
and are therefore not reproduced in figs.~1 and 2.
They are generally correlated with the bounds on chargino masses
\cite{chino}.

\section{The lighter $CP$-even Higgs boson $h^0$}
\label{sec:Xsects-h}
\setcounter{equation}{0}
Let us first consider the cross section for
\beq
pp\to h^0 X
\eeq
For $\MNS=200~\GeV$, $\Ms=500~\GeV$, and $\mu=500~\GeV$,
we show in fig.~3 this cross section for four values of $A$,
the trilinear coupling parameter.
The following features are rather striking:
\begin{itemize}
\item The cross section decreases appreciably for large values of $A$.
This is mainly due to an increase in the $h^0$ mass.
\item There are sharp edges at small values of $\tan\beta$, and
also at small $m_A$.
The edges at small $\tan\beta$ are caused by the $h^0$ becoming light.
At small values of $m_A$ and large $A$, the couplings of $h^0$
to $b$ quarks and $\tau$ leptons become large, making
the cross section very large in this region.
\end{itemize}

For the same parameters as above, we show in fig.~4
the total decay rate, $\Gamma(h^0\to\mbox{all})$ and the
two-photon decay rate, $\Gamma(h^0\to\gamma\gamma)$.
As opposed to fig.~3, here we only consider two values of $A$,
namely $A=0$ and $A=1000~\GeV$.
The two-photon decay rate is seen to increase sharply at large
values of $A$, but this does not result in a larger rate for
the process
\beq
\label{eq:pp-h0-gammagamma}
pp\to h^0 X \to \gamma\gamma X
\eeq
since the production cross section also decreases, as shown in fig.~3
(mostly due to an increase in the Higgs mass, $m_{h^0}$).
In fig.~5 we show the cross section for the process
(\ref{eq:pp-h0-gammagamma}).
A characteristic feature of the cross section is that it is small
at moderate values of $m_A$, and then increases steadily with
increasing $m_A$, reaching asymptotically a plateau.
This behaviour is caused by the contribution of the $W$ to the
triangle graph for $h^0\to\gamma\gamma$.
The $h^0WW$ coupling is proportional to $\sin(\beta-\alpha)$,
where $\alpha$ is defined in terms of masses
(including radiative corrections) as
\beq
\cos2\alpha=-\cos2\beta\left({m_{A}^2-m_Z^2\over m_{H^0}^2-m_{h^0}^2}\right),
\qquad -{\pi\over2}\le\alpha\le0.
\eeq
For large $m_A$, at fixed $\beta$, all Higgs masses, except $m_{h^0}$,
become large, so that $h^0$ decouples. For large $m_{A}$, we
actually have $\sin(\beta-\alpha)\to1$,
which is why the cross section increases and reaches a plateau
for large $m_{A}$.

In figures 6 and 7 we show contour plots of the cross section
(\ref{eq:pp-h0-gammagamma}), for $\MNS=200~\GeV$, $\Ms=500~\GeV$
and $\mu$ equal to $-500$ and $+500$~GeV, respectively.
For each case, four values of the trilinear chiral-mixing parameter
$A$ are considered, $A=0$, 200~GeV, 500~GeV and 1000~GeV
(figure~7a is thus a different representation of figure~5).

The $\mu$-dependence of the cross section can for the case of
$\MNS=200~\GeV$ and $\Ms=500~\GeV$ be described as follows.
At moderate values ($\mu=\pm200~\GeV$), there is not much difference
between the cross section for positive and negative values of $\mu$.
The cross section has a significant dependence on $m_A$,
being low at $m_A\le\Order(300~\GeV)$,
then increasing steadily and reaching a plateau with increasing $m_A$.
The dependence on $\tan\beta$ is rather weak.

For increasing values of $|\mu|$ (500~GeV, 1000~GeV)
the change in the cross section is rather complex.
This is basically caused by two phenomena:
(1) At large values of $|\mu|$ the squarks become too light
or unphysical, in analogy with the case ($\MNS=200~\GeV$,
$\Ms=150~\GeV$) shown in fig.~1.
Hence, there are regions both at small and large values of $\tan\beta$
where the cross section is not defined.
(2) At large values of $|\mu|$ and large values of $\tan\beta$
(all $m_A$) the Higgs gets very light (due to radiative
corrections). As a consequence, the cross section can get rather
high, where not forbidden due to (1) above.

The dependence of the cross section on $\MNS$ and $\Ms$ is
described in table~1.
For small values of $\Ms$ ($\sim150~\GeV$), the possible ranges of
$\tan\beta$, $\mu$ and $A$ become severely restricted, in order to
obtain physically acceptable squark masses.
There is a significant increase in the cross section as $\Ms$
increases from 500~GeV to 1~TeV, to values of the order of 25--30~fb.

For small values of $\MNS$ ($\sim50~\GeV$), the possible range in $\mu$
is restricted in order to obtain physically acceptable chargino masses.
As $\MNS$ increases beyond 200~GeV,
there is little further change in the cross section.
Details are given in table~1.

The cross section has a modest dependence on the choice of
gluon distribution function used.
For the plots shown here, we have used the recent GRV Set~3 \cite{GRV}
distributions, which are the default of the PDFLIB.
The BM Set~1 \cite{BM} leads to an increase of the cross section
by about 5--6\%, whereas the MRS Set~29 (S0') \cite{MRS} and CTEQ Set~24
(2pM) \cite{CTEQ} give reductions by 3--5\% and 8--9\%, respectively.
These uncertainties are thus rather insignificant.
\section{The heavier $CP$-even Higgs boson $H^0$}
\label{sec:Xsects-H}
\setcounter{equation}{0}
We shall here briefly consider the process
\beq
\label{eq:pp-H0-gammagamma}
pp\to H^0 X \to \gamma\gamma X
\eeq
which is of interest for small values of $m_A$.

The two-photon decay of the heavier $CP$-even Higgs
proceeds dominantly through $W$ loops, and is
complementary to that of the lighter $CP$-even Higgs.
It is only significant if $m_A$ is small,
hence $m_{H^0}$ itself must also be light.
At small $m_A$ the total $H^0$ decay rate is small
and thus the branching ratio for it to go into
two photons can be considerable.
As a result, the cross section for the process
(\ref{eq:pp-H0-gammagamma}) can at small values of $m_A$
reach values exceeding 200~fb.

Contour plots of the cross section are shown in
figures~8 and 9, for four values of $A$, and for $\mu=-500~\GeV$
(fig.~8) and $\mu=500~\GeV$ (fig.~9).
For $\mu=-500~\GeV$ there is a strong {\it increase} in the cross
section with increasing values of $A$.
For positive values of $\mu$, however, increasing values
of $A$ lead to a reduction of the cross section.
At low values of $m_A$ the Higgs mass $m_{H^0}$
is of the order of 110--140~GeV, and the $h^0$
mass is close to the experimental lower limit.
\section{Summary and concluding remarks}
\label{sec:conc}
\setcounter{equation}{0}
We have discussed in some detail the cross section for
producing the $CP$-even Higgs bosons at the LHC,
in conjunction with their decay to two photons.
Where the parameters lead to a physically acceptable phenomenology,
the cross section multiplied by the two-photon branching ratio
is for the lighter $CP$-even Higgs boson of the order of 20--30~fb.

These numbers do not take into account QCD corrections.
Such corrections have been evaluated for the quark-loop contribution,
and lead to enhancements of the cross section of about 50\% \cite{Spira}.
However, in the presence of chiral mixing the squark loops also
contribute significantly. Since the QCD corrections for these
are not available, we have decided it was more clean to simply
leave out all higher-order QCD effects.
One should of course keep in mind that they are very important.

There is a modest increase of the cross section with increasing
values of $A$ (i.e., with increasing chiral mixing).
This comes about as the result of two competing effects:
with increasing $A$, the Higgs boson becomes more heavy,
leading to lower production cross sections.
This is however offset by a corresponding increase
in the two-photon decay rate.

This research has been supported by the Research Council of Norway.
The work of PNP was supported by the Department of Science
and Technology under project No. SP/S2/K-17/94.
One of the authors (PNP) would like to thank NORDITA and the
University of Bergen for hospitality while this work was done.
We are also grateful to P. Janot, P. Zerwas and F. Zwirner
for useful correspondence and interest in this work.

\renewcommand{\arraystretch}{1.5}  
\clearpage
{\textwidth 20cm 
Table~1.
Dependence of the cross section for $pp\to h^0X\to\gamma\gamma X$
on $\MNS$, $\Ms$ and $\mu$
\medskip

\begin{tabular}{||r||c|c|c||} \hline
$\MNS$\ \ \ \ \  &
\begin{describe}
\begin{center}
$\Ms=150~\GeV$
\end{center} \vspace{-2mm}
lightest squark too light for large values of $\tan\beta$
and increasing $A$
\end{describe}
& $\Ms=500~\GeV$ & $\Ms=1000~\GeV$ \\
\hline \hline
\begin{describea}
\flushright 50~GeV \\ \vspace{2mm}
lightest chargino too light for small values of $\tan\beta$
and positive $\mu$
\end{describea} &
\begin{describe}
cross section significant only for small $A$, small $\tan\beta$,
and small negative $\mu$
\end{describe} &
\begin{describe}
cross section lower than at $\MNS=200~\GeV$, especially at moderate,
positive $\mu$;
for large $|\mu|$
cross section significant only for
narrow range in $\tan\beta$
\end{describe} &
\begin{describe}
cross section larger than at $\Ms=500~\GeV$
\end{describe} \\ \hline
200~GeV &
\begin{describe}
cross section significant only for small $A$, small $\tan\beta$,
and small $|\mu|$
\end{describe} &
\begin{describe}
``default" given in figs.~3--9 for $\mu=\pm500~\GeV$;
complex dependence on $\tan\beta$ for larger $|\mu|$
\end{describe} &
\begin{describe}
cross section significantly {\em larger} than at $\Ms=500~\GeV$,
reaching well beyond 25~fb for large range of $\mu$;
less dependence on $\tan\beta$ than at $\Ms=500~\GeV$,
in particular for large $A$
\end{describe} \\ \hline
\begin{describea}
\flushright 500~GeV \\ \vspace{1mm}
\flushright 1000~GeV
\end{describea} &
\begin{describe}
very similar to $\MNS=200~\GeV$
\end{describe} &
\begin{describe}
very similar to $\MNS=200~\GeV$
\end{describe} &
\begin{describe}
very similar to $\MNS=200~\GeV$
\end{describe} \\ \hline
\end{tabular}
} 

\clearpage
\centerline{\bf Figure captions}

\vskip 15pt
\def\fig#1#2{\hangindent=.65truein \noindent \hbox to .65truein{Fig.\ #1.
\hfil}#2\vskip 2pt}

\fig{1}{Regions in the $\mu$-$\tan\beta$ plane which are ruled out
by too light chargino ($\chi^\pm$) and squark masses.
The gaugino mass scale is $\MNS=200~\GeV$, $\Ms=150~\GeV$,
and $m_A=200~\GeV$.
At this low value of $\Ms$, the squark masses are too light or
unphysical in much of the $\mu$-$\tan\beta$ plane.
The hyperbola-like contours give regions that are excluded
by the lightest $b$ squark being below 45~GeV (solid) or
90~GeV (dashed).
The more straight contours at large $\mu$ and small $\tan\beta$
similarly indicate regions that are excluded by the
lightest $t$ squark.
In a) we consider the trilinear mixing parameter $A=0$, whereas
in b) we take $A=200~\GeV$.
}

\fig{2}{Regions in the $\mu$-$\tan\beta$ plane which are ruled out
by too light chargino ($\chi^\pm$) and $h^0$ masses.
Similar to fig.~1, but for $\Ms=500~\GeV$.
In a) we consider the trilinear mixing parameter $A=0$, whereas
in b) we take $A=1000~\GeV$.
The solid (dashed) contours for small $|\mu|$ refer to the chargino mass
$m_{\chi^\pm}=45$ (90)~GeV.
The unlabelled contours near the corners at large $|\mu|$
refer to regions where the $h^0$ mass would be below 45~GeV.
}

\fig{3}{Cross section for $pp\to h^0X$
as a function of $m_A$ and $\tan\beta$
for $\MNS=200~\GeV$,
$\Ms=500~\GeV$, and $\mu=500~\GeV$.
Four values of $A$ are considered: a)~$A=0$,
b)~$A=200~\GeV$, c)~$A=500~\GeV$ and d)~$A=1000~\GeV$.
}

\fig{4}{Total decay rate $\Gamma(h^0\to\mbox{all})$ and
two-photon decay rate $\Gamma(h^0\to\gamma\gamma)$,
as functions  of $m_A$ and $\tan\beta$
for $\MNS=200~\GeV$,
$\Ms=500~\GeV$, and $\mu=500~\GeV$.
Two values of $A$ are considered: $A=0$
and $A=1000~\GeV$.
}

\fig{5}{Cross section for $pp\to h^0X\to\gamma\gamma X$
as a function of $m_A$ and $\tan\beta$
for $\MNS=200~\GeV$,
$\Ms=500~\GeV$, $\mu=500~\GeV$, and $A=0$.
}

\fig{6}{Dependence of the $pp\to h^0\to \gamma\gamma$
cross section on $m_A$ and $\tan\beta$
for different values of the trilinear couplings $A$.
Four values of $A$ are considered: a)~$A=0$,
b)~$A=200~\GeV$, c)~$A=500~\GeV$ and d)~$A=1000~\GeV$.
Here $\MNS=200~\GeV$, $\Ms=500~\GeV$, and $\mu=-500~\GeV$.
The solid contours are at 15~fb, the long-dashed ones at 20~fb,
and the short-dashed ones at 25~fb.}

\fig{7}{Dependence of the $pp\to h^0\to \gamma\gamma$
cross section on $m_A$ and $\tan\beta$
for different values of the trilinear couplings $A$.
As fig.~6, except that $\mu=500~\GeV$.}

\fig{8}{Dependence of the $pp\to H^0\to \gamma\gamma$
cross section on $m_A$ and $\tan\beta$
for different values of the trilinear couplings $A$.
Four values of $A$ are considered: a)~$A=0$,
b)~$A=200~\GeV$, c)~$A=500~\GeV$ and d)~$A=1000~\GeV$.
Here $\MNS=200~\GeV$, $\Ms=500~\GeV$, and $\mu=-500~\GeV$.
The contours are at 10~fb (solid), 20~fb, 50~fb, 100~fb and 200~fb.}

\fig{9}{Dependence of the $pp\to H^0\to \gamma\gamma$
cross section on $m_A$ and $\tan\beta$
for different values of the trilinear couplings $A$.
As in fig.~8, except that $\mu=500~\GeV$.}




\begin{figure}
\refstepcounter{figure}
\begin{center}
\mbox{\epsffile{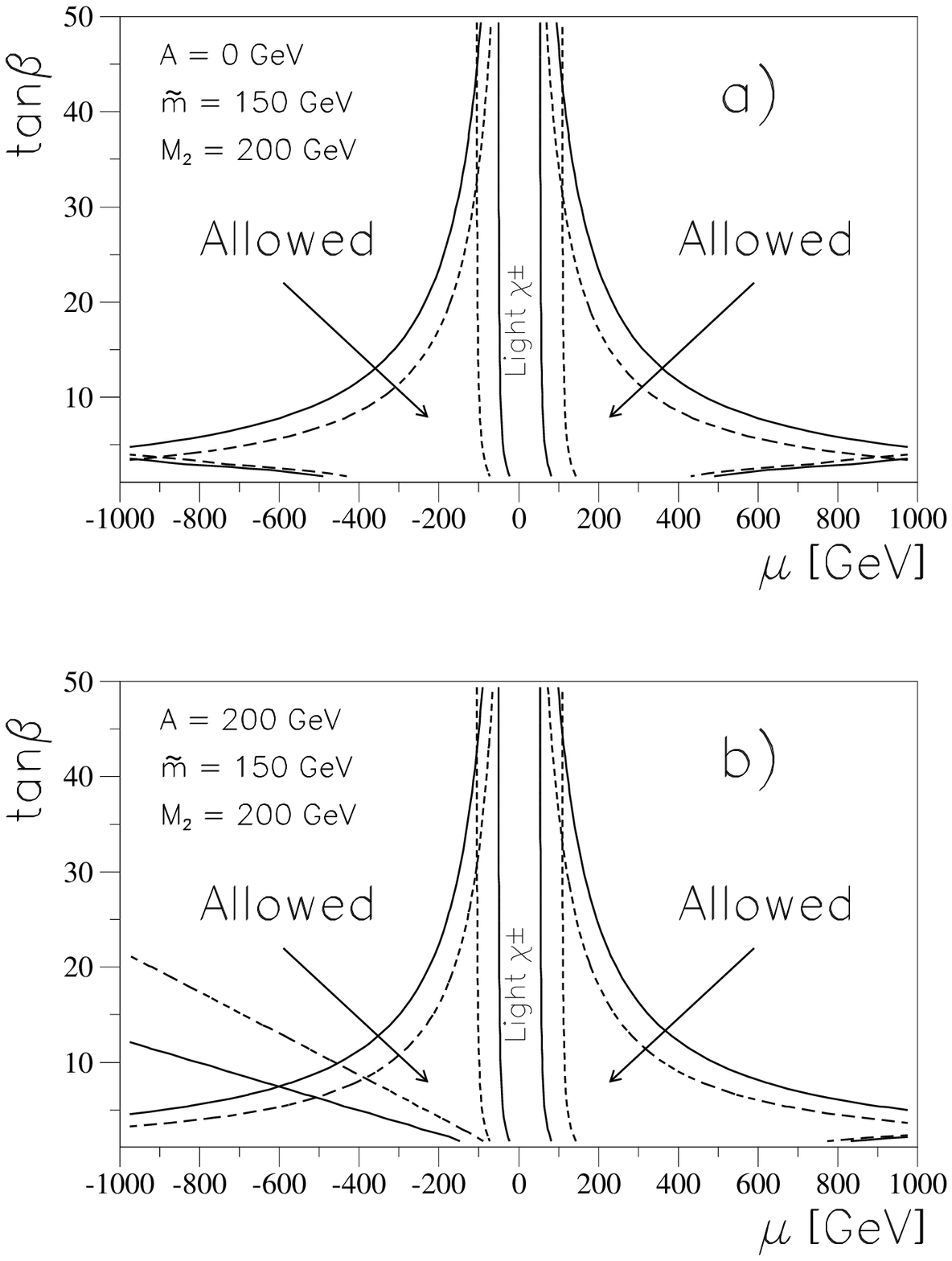}}
\vspace{20mm}
Figure~\thefigure
\end{center}
\end{figure}


\begin{figure}
\refstepcounter{figure}
\begin{center}
\mbox{\epsffile{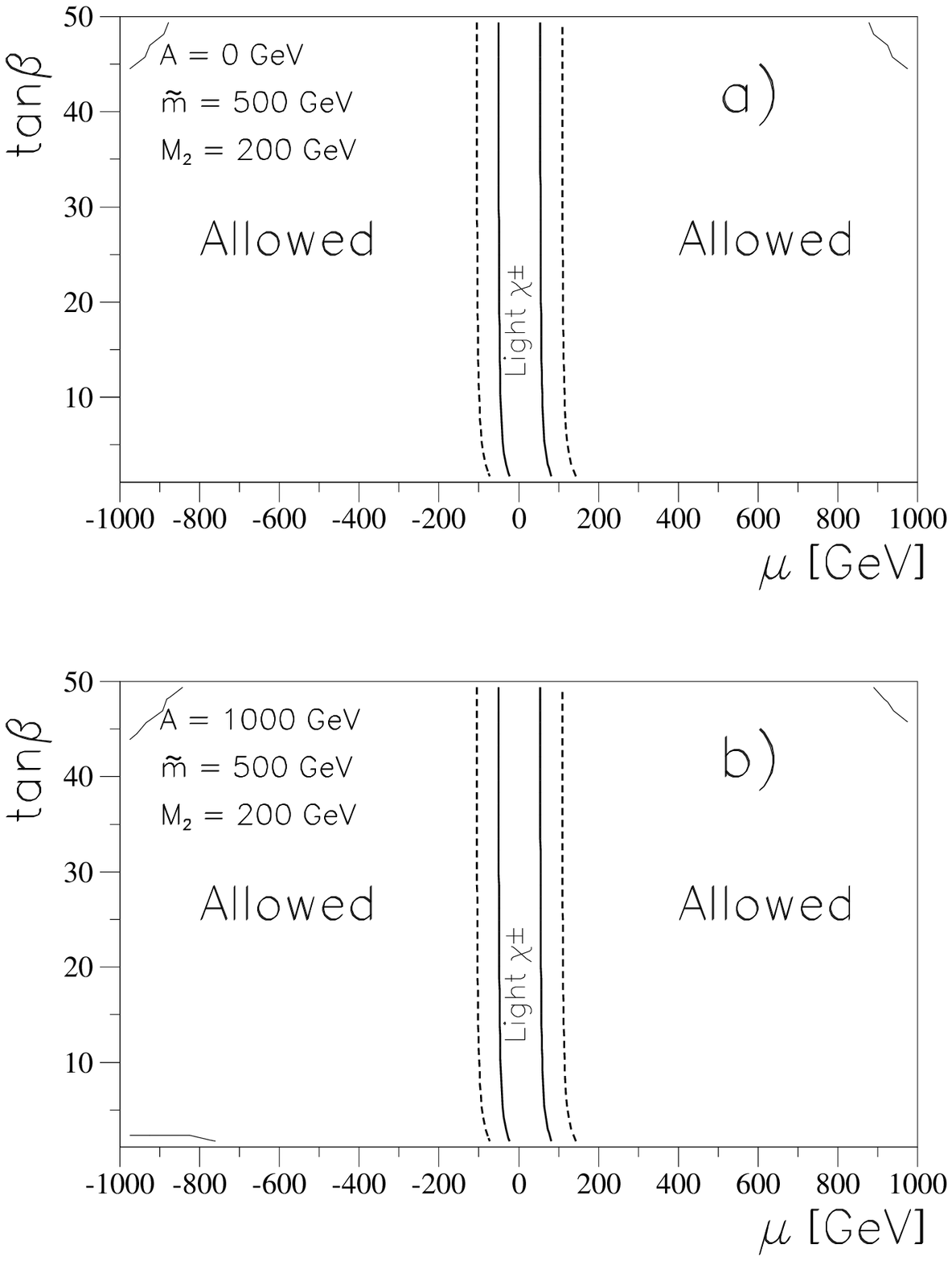}}
\vspace{20mm}
Figure~\thefigure
\end{center}
\end{figure}


\begin{figure}
\refstepcounter{figure}
\begin{center}
\mbox{\epsffile{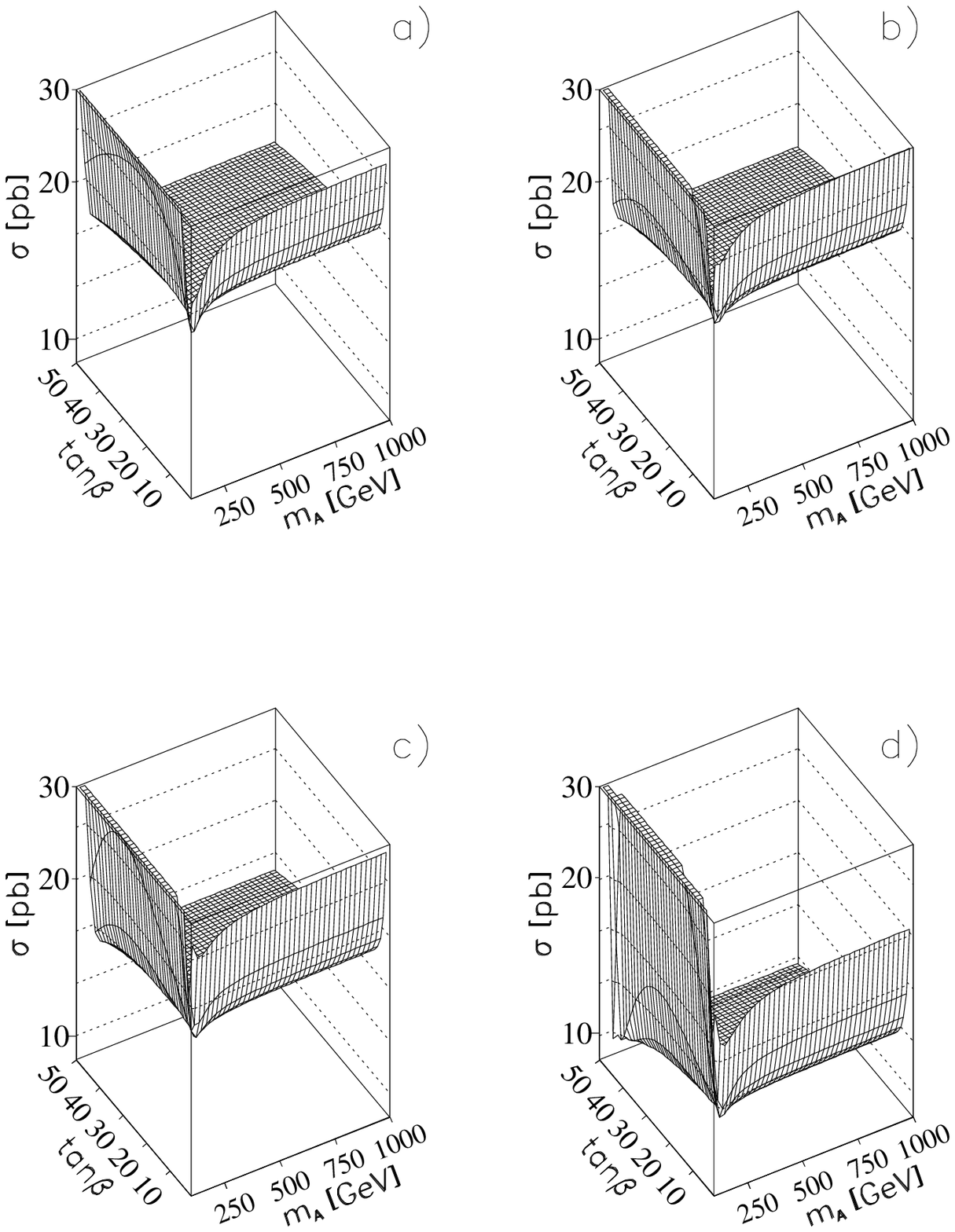}}
\vspace{20mm}
Figure~\thefigure
\end{center}
\end{figure}

\begin{figure}
\refstepcounter{figure}
\begin{center}
\mbox{\epsffile{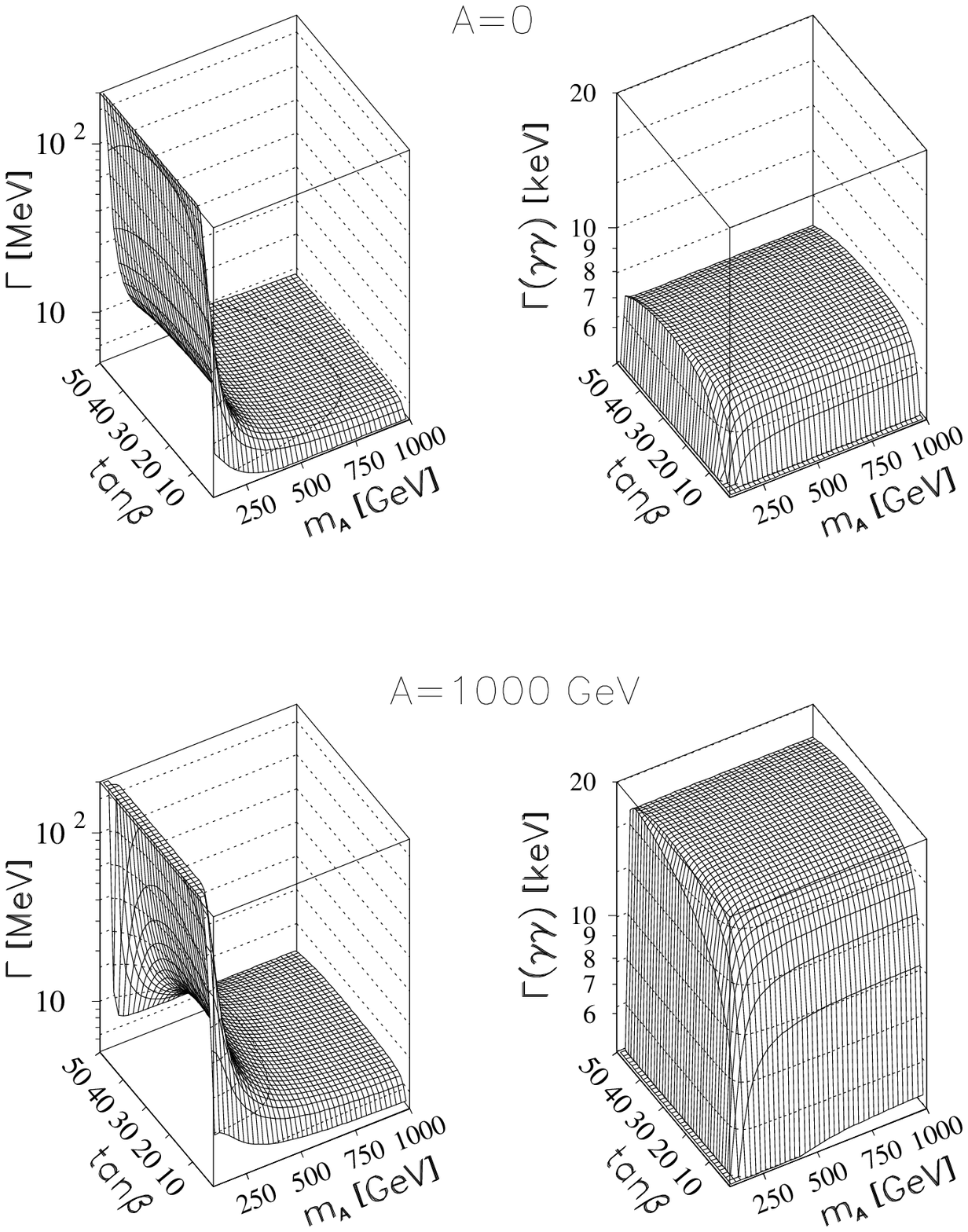}}
\vspace{20mm}
Figure~\thefigure
\end{center}
\end{figure}

\begin{figure}
\refstepcounter{figure}
\begin{center}
\mbox{\epsffile{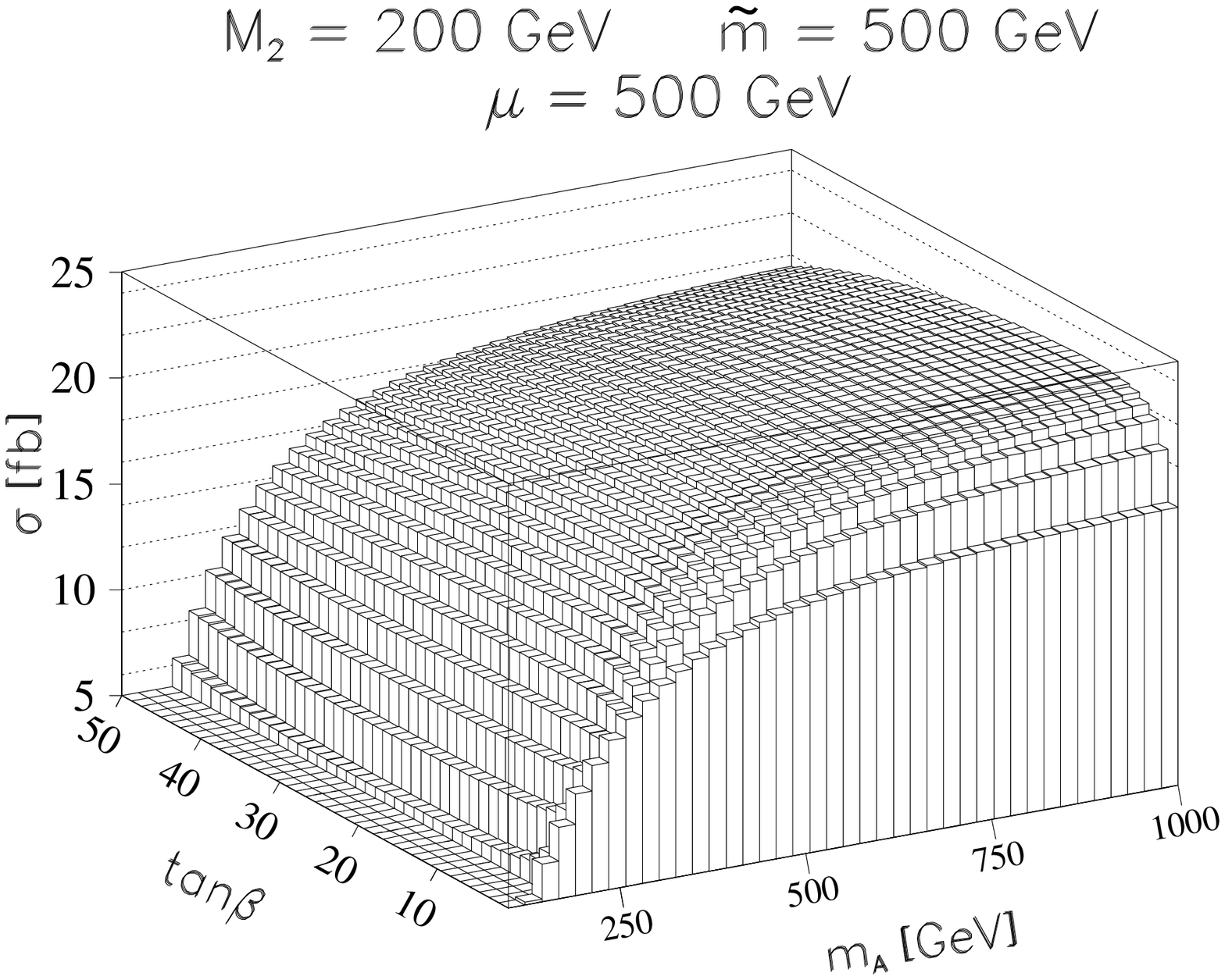}}
\vspace{20mm}
Figure~\thefigure
\end{center}
\end{figure}

\begin{figure}
\refstepcounter{figure}
\begin{center}
\mbox{\epsffile{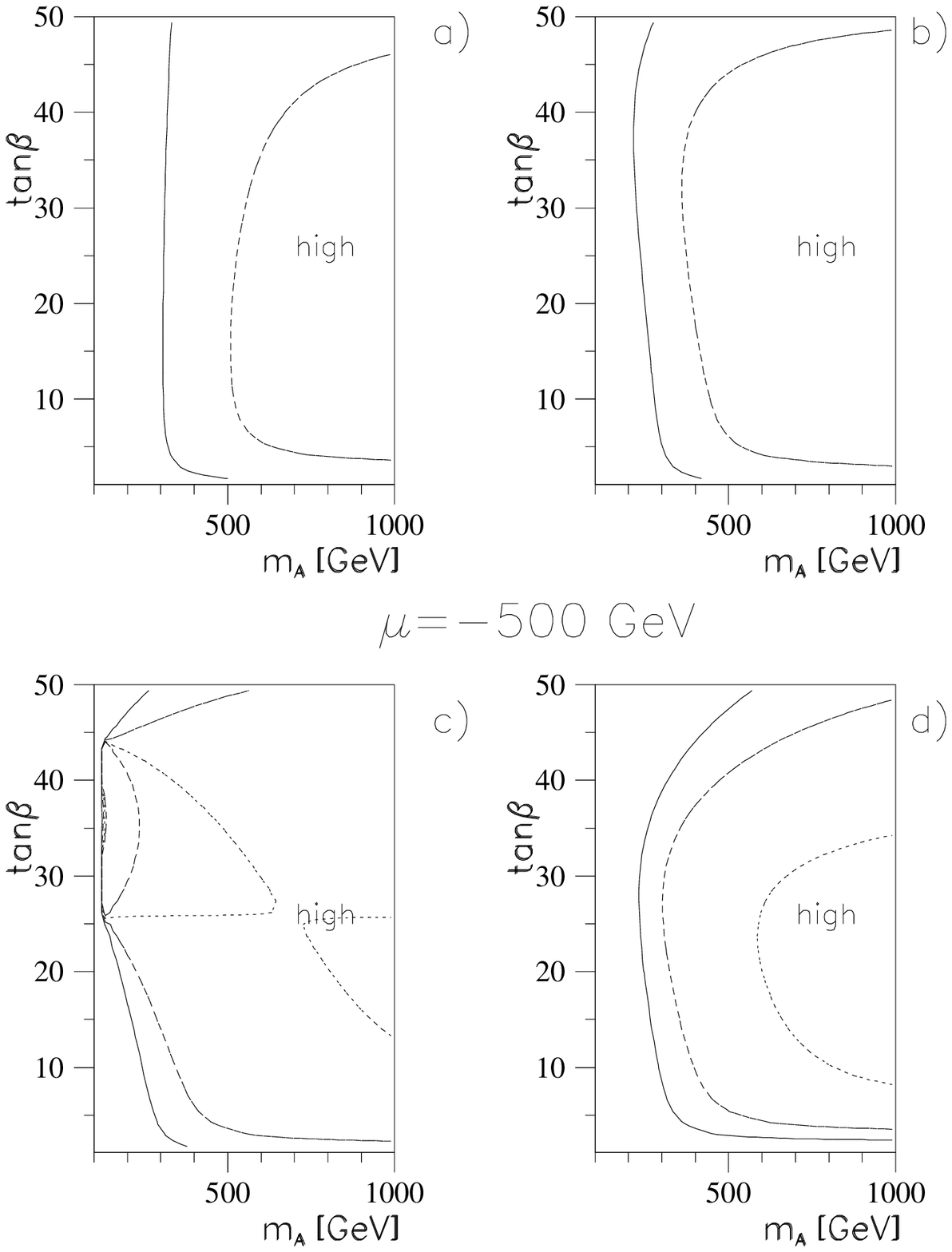}}
\vspace{20mm}
Figure~\thefigure
\end{center}
\end{figure}

\begin{figure}
\refstepcounter{figure}
\begin{center}
\mbox{\epsffile{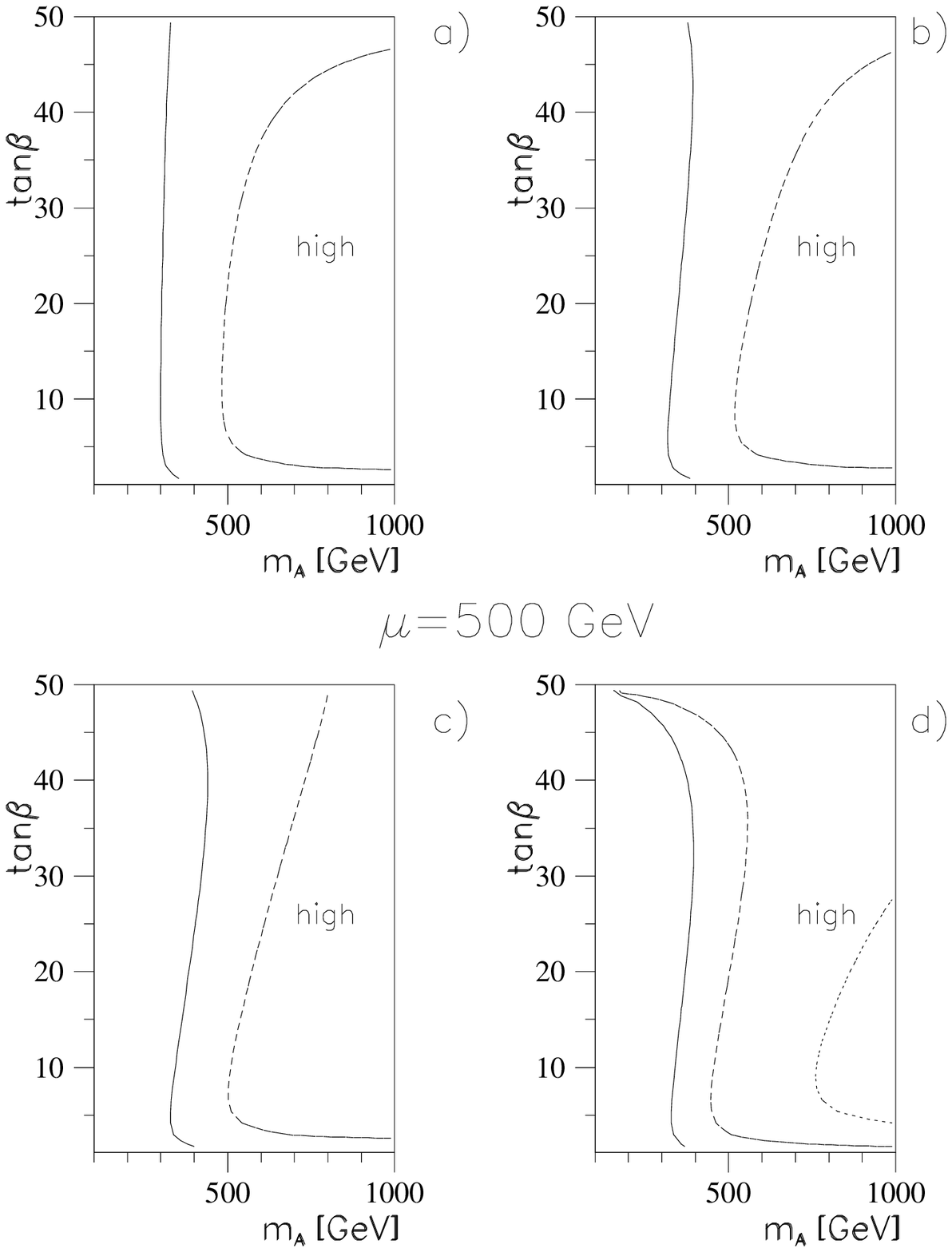}}
\vspace{20mm}
Figure~\thefigure
\end{center}
\end{figure}

\begin{figure}
\refstepcounter{figure}
\begin{center}
\mbox{\epsffile{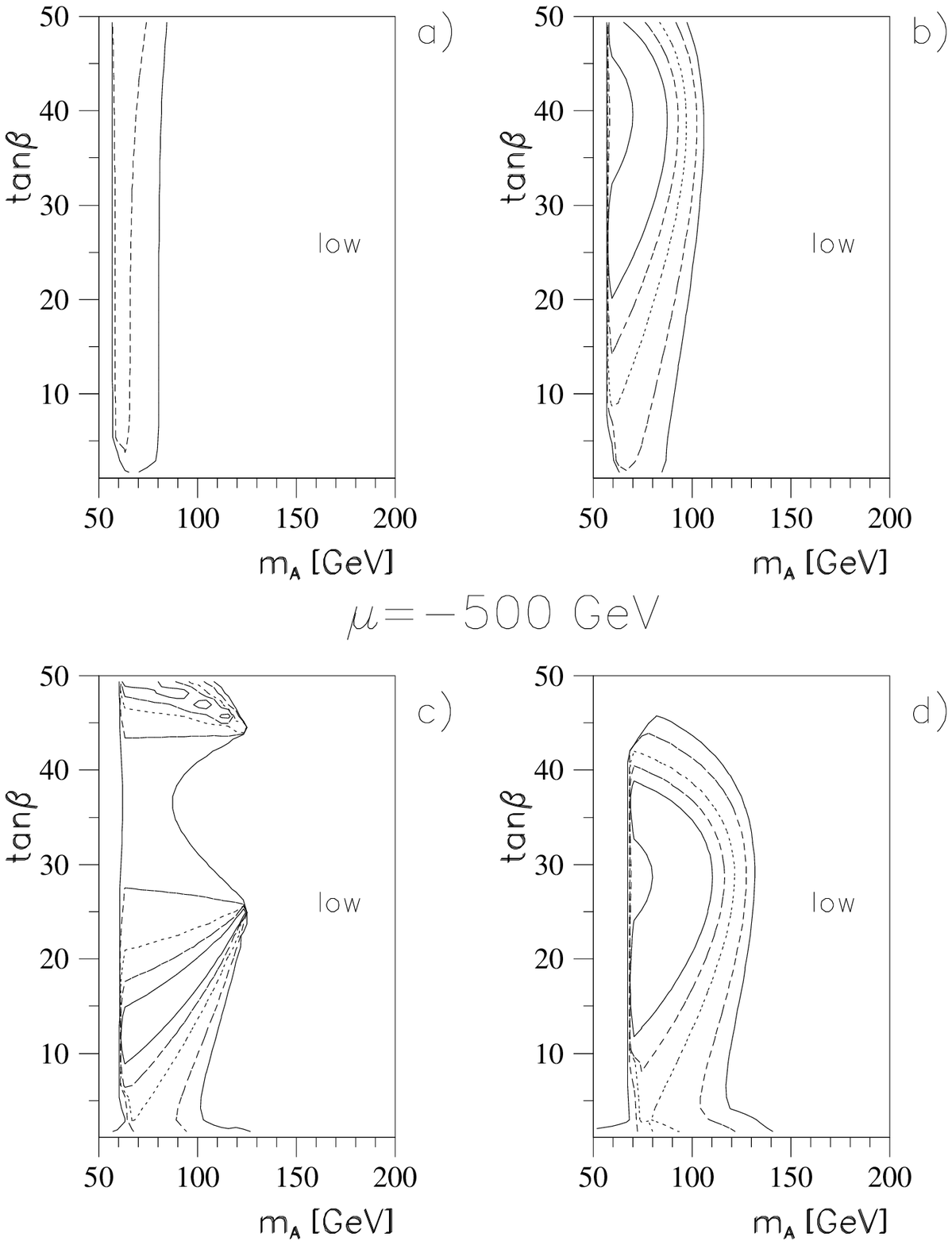}}
\vspace{20mm}
Figure~\thefigure
\end{center}
\end{figure}

\begin{figure}
\refstepcounter{figure}
\begin{center}
\mbox{\epsffile{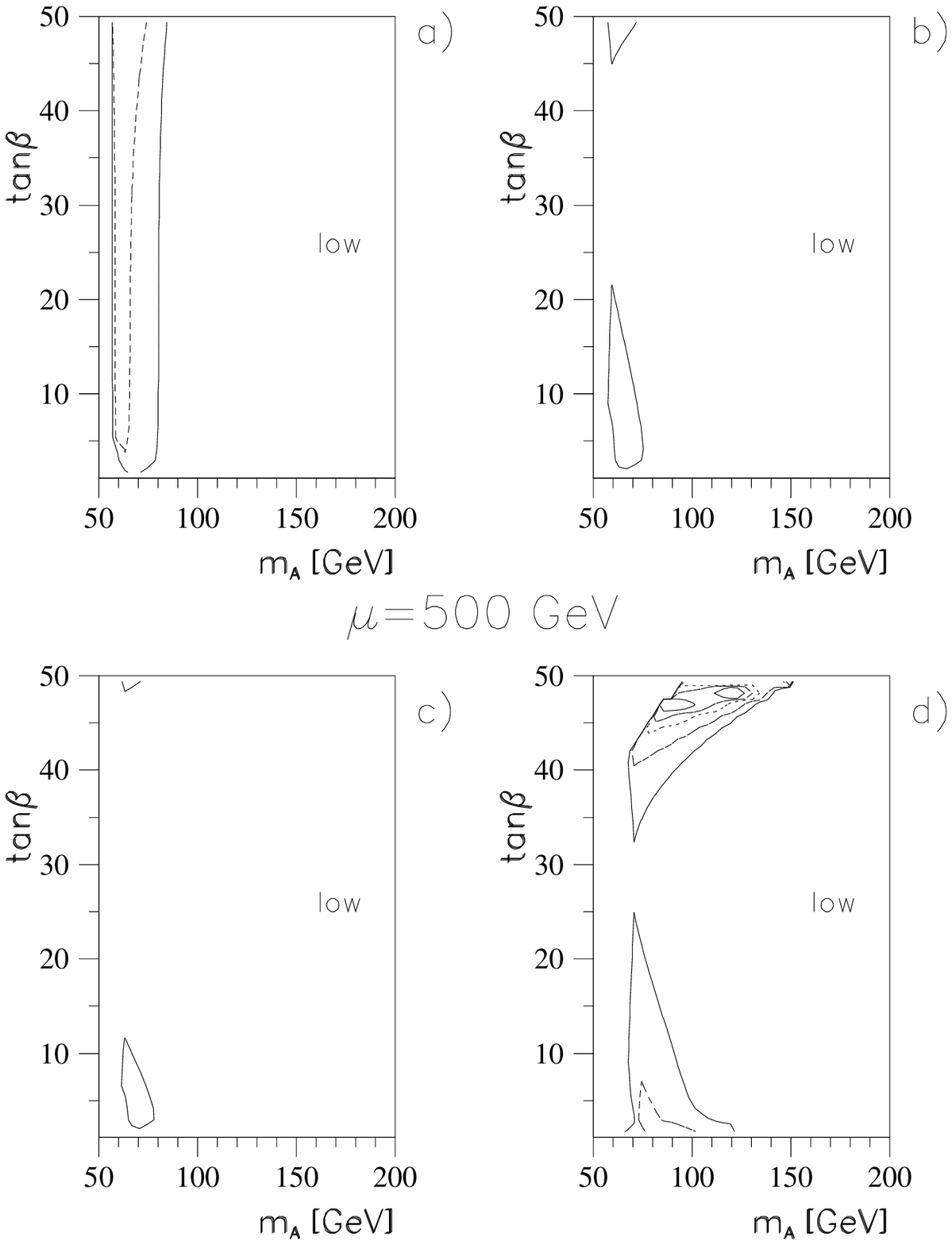}}
\vspace{20mm}
Figure~\thefigure
\end{center}
\end{figure}

\end{document}